\documentclass[%
 reprint,
superscriptaddress,
 bibnotes,
 amsmath,
 amssymb,
 aps,
pra,
]{revtex4-2}

\usepackage{graphicx}
\usepackage{dcolumn}
\usepackage{bm}
\usepackage{physics}
\usepackage[inkscapelatex=false]{svg}
\usepackage{bbm}
\usepackage{mathrsfs}
\usepackage{xcolor}
\usepackage[most]{tcolorbox}

\newtcolorbox{highlighted}{colback=yellow,coltext=red,breakable}
\begin{document}

\preprint{APS/123-QED}

\title{Qualitatively altered driven Dicke superradiance in extended systems due to infinitesimal perturbations}

\author{Wenqi Tong}
 \affiliation{Elmore Family School of Electrical and Computer Engineering, Purdue University, West Lafayette, Indiana 47907, USA}
\author{F. Robicheaux}%
 \email{robichf@purdue.edu}
 \affiliation{%
 Department of Physics and Astronomy, Purdue University, West Lafayette, Indiana 47907, USA
}%
 \affiliation{Purdue Quantum Science and Engineering Institute, Purdue University, West Lafayette, Indiana 47907, USA}




\date{\today}

\begin{abstract}
The driven Dicke model, with interesting quantum phases induced by parameterized driving, has been intensively studied in cavities, where permutation symmetry applies due to the atoms' equal coupling to the field and identical interaction. As a result, the system, with proper initialization, can remain in a highly symmetric subset of the state space, where the photon emission of each atom constructively interferes with each other, leading to superradiance at steady state. However, because of the degeneracy of steady states for the driven Dicke model, the steady state can be qualitatively altered by an infinitesimal perturbation. In this work, we simulate superconducting qubits coupled to 
 a 1D waveguide as the extended system and theoretically investigate four kinds of perturbations: local dephasing, individual driving phases, the separation between adjacent qubits, and individual detunings. Using an angular momentum basis, we predict the dimension of the degenerate subspace and study the transition within the subspace due to the perturbation.
\end{abstract}

\maketitle

\section{\label{sec:introduction}Introduction}


Dicke superradiance\cite{PhysRev.93.99,PhysRevA.3.1735,gross1982superradiance}, the collective radiation of $N$ inverted atoms, characterized by a burst of light with peak intensity scaled by $N^2$, instead of $N$, arises from the spontaneous synchronization of the atoms during the decay, with phase locked and photon emission rate enhanced. Such phenomenon, sometimes also called ``superradiant burst'', has been observed in a variety of platforms\cite{PhysRevLett.30.309,PhysRevLett.36.1035,PhysRevLett.49.117,scheibner2007superradiance,rohlsberger2010collective,inouye1999superradiant,PhysRevLett.98.053603,PhysRevLett.76.2049,eschner2001light} and analytically solved in the thermodynamic limit\cite{PhysRevA.106.013716}. In recent years,  particular attention has been devoted to superradiance due to its potential to develop lasers with ultranarrow linewidth\cite{PhysRevLett.102.163601,Maier:14,bohnet2012steady,PhysRevX.6.011025,norcia2016superradiance,PhysRevLett.123.103601,PhysRevA.101.013819}.


In addition to the Dicke superradiance in an initially inverted population, driving the system introduces another degree of freedom and the interplay between driving and dissipation gives rise to many novel phases\cite{PhysRevLett.114.040402,PhysRevA.89.023616,PhysRevA.97.053616,PhysRevLett.110.257204,PhysRevResearch.3.L022020,parmee2020signatures,PhysRevA.95.053833,PhysRevLett.107.113602,PhysRevA.96.053857,PhysRevLett.117.173601}. To overcome the exponential growth of the Hilbert space dimension, the conventional Dicke model\cite{PhysRev.93.99,PhysRevA.3.1735,gross1982superradiance} or driven Dicke model\cite{PhysRevA.98.042113,PhysRevLett.110.080502,PhysRevA.108.023725} assumes that the atoms are confined in a small spatial volume or specifically placed so that the atoms are considered indistinguishable with identical interaction between them. As a result, the system remains in a highly symmetric subspace $\ket{S = N/2, M}$ if initialized in it, whose dimension is linear in the atom number, $N$, so that the accurate numerical simulation of the system is feasible, even for large $N$. 


In recent years, the development of ordered atomic arrays, featuring almost arbitrary manipulation of atoms' locations\cite{kim2016situ,endres2016atom,barredo2016atom,norcia2018microscopic,PhysRevLett.122.143002,PhysRevLett.122.203601,bakr2010probing,sherson2010single,greif2016site,kumar2018sorting}, opens a new possibility to unveil intriguing but complicated many-body physics beyond the Dicke limit\cite{bohnet2012steady,PhysRevA.105.053715,PhysRevLett.130.053601,parmee2020signatures,PhysRevLett.110.080502}. Extended systems, with atomic separation greater than the wavelength of the emitted light, have been investigated for Dicke superradiance in \cite{masson2022universality,PhysRevLett.130.213605,PhysRevResearch.4.023207,PRXQuantum.5.010344}. In addition, artificial atoms realized as superconducting qubits, with convenience to address each qubit\cite{ma2019dissipatively,wallraff2004strong}, 
also exhibit Dicke superradiance\cite{mlynek2014observation,PhysRevB.94.224510}. Especially, when coupled to a 1D waveguide, where the field propagates with negligible damping, the superconducting qubits feature long-range interaction\cite{van2013photon,PhysRevA.88.043806} and can be a good candidate for the extended system under investigation in this work.


Recent experiments report the transition to Dicke superradiance using a cloud of atoms driven by an external light field in free space\cite{ferioli2023non,PhysRevLett.127.243602}, followed by the theoretical explanation in \cite{goncalves2024driven}. The driven Dicke superradiance in extended systems emerges as an interesting topic and motivates the discussion of how the geometry of the system \cite{PhysRevResearch.6.023206} or initial states\cite{PhysRevA.109.032204} alters the steady state. There are also discussions of other factors like dephasing \cite{PhysRevLett.118.123602,PhysRevA.94.061802,PhysRevA.96.023863}, individual driving phase\cite{PhysRevA.109.013715,PhysRevA.108.023708,PhysRevA.106.053703}, and detuning\cite{PhysRevLett.95.243602,stannigel2012driven,ramos2014quantum,PhysRevA.91.042116,lei2023many}. However, their effects on the steady states of driven-dissipative systems are not fully explored. 

We focus on transmons\cite{9789946}, a kind of superconducting qubit with less sensitivity to charge noise, and investigate the effect of four kinds of perturbations (dephasing, changing the separation, varying the driving phases, and inhomogeneous detunings) on the steady state of a transmon lattice coupled to a 1D waveguide. Simulations show that even an infinitesimal perturbation can significantly change the steady state. Diagonalization of the Liouvillian matrix reveals a degenerate subspace of possible steady states when the perturbation is exactly $0$. Although the degeneracy and perturbation theory of Liouvillian has been discussed in \cite{li2014perturbative,huybrechts2024quantum,thingna2021degenerated,albert1802lindbladians,gomez2018perturbation,li2016nonequilibrium,krishna2023select}, as far as we know, the perturbation theory on the degenerate state space and its application to explain the imperfect driven Dicke model has not yet been explored. In addition, we use an angular momentum basis to count the number of degenerate states. This basis can be used to reproduce parts of the eigenvalues using first-order perturbation theory in the degenerate subspace. 

This work is organized as follows: Section \ref{sec:theory} describes the basic equations of motion, the expressions of perturbation, and our observation. Section \ref{sec:results} exhibits the time evolution of the observables, illustrating the significant change in steady states for an infinitesimal perturbation. In Sec. \ref{sec:explanation} we explain the phenomena using the diagonalization of the Liouvillian matrix and develop an angular momentum scheme to explain the degeneracy and the eigenvalues.

\section{\label{sec:theory}Theory}
\subsection{Equations of motion, perfect driven Dicke Model}
To investigate the effect of perturbations on the perfect driven Dicke Model, it is necessary to study the dynamics of the unperturbed system. In this work, we use transmons coupled to a 1D waveguide to build up the extended system and model the light by classical continuous wave driving. Because the energy gap between the lowest two levels and the anharmonicity are much greater than the linewidth, the transmons are approximated by two-level systems\cite{blais2020quantum}. Throughout the paper, we use $\ket{0}, \ket{1}$ to represent the ground state and first excited state of a transmon. The operators involved in this work are defined as below:
\begin{equation}
    \hat{\sigma}^+_n \! = \! \ket{1}_n\!\bra{0}_n,  \hat{\sigma}^-_n \! = \! \ket{0}_n\!\bra{1}_n,  \hat{\sigma}^z_n \! = \! \ket{0}_n\!\bra{0}_n \! - \! \ket{1}_n\!\bra{1}_n
\end{equation}
We also make the Markovian approximation and trace out the photon degrees of freedom, assuming that the time scale of photon propagation is much smaller than the interaction\cite{PhysRevA.2.883,PhysRevA.48.3365,PhysRevA.52.636,FICEK2002369,PhysRevA.72.063815,PhysRevB.82.075427,PhysRevLett.106.020501,PhysRevB.86.024503,Chang_2012,PhysRevLett.110.113601}. By the Green's function formalism, the equation of motion follows the form \cite{PhysRevA.95.033818,PhysRevResearch.2.043213,PhysRevResearch.3.033233}:
\begin{equation}
    \begin{split}
        \frac{d\hat{\rho}}{dt} = & - i [H, \hat{\rho}] + \mathcal{L} [\hat{\rho}], 
    \end{split}
    \label{eq:eom}
\end{equation}
where we set $\hbar = 1$. The Hamiltonian, after the rotating wave approximation, includes two parts:

\begin{equation}
    H = H_l + H_{wg}
\end{equation}
where $H_l$ describes the coherent driving:
\begin{equation}
    H_l = \sum_n \left (\frac{\Omega_n}{2} \hat{\sigma}_n^+ + \frac{\Omega_n^*}{2} \hat{\sigma}_n^- - \frac{\Delta_n}{2} \hat{\sigma}_n^z \right ),
    \label{eq:H_l}
\end{equation}
where $\Omega_n$ is the Rabi frequency of the $n$th transmon and carries the phase $\phi_n$:
\begin{equation}
    \Omega_n = \Omega e^{i\phi_n},
\end{equation}
and $\Delta_n$ is the detuning of the $n$th transmon.

$H_{wg}$ refers to the transmon-transmon exchange interaction through the waveguide, with the strength denoted by $\Omega_{nm}$:
\begin{equation}
    H_{wg} = \sum_{n, m} \Omega_{nm} \hat{\sigma}_m^+ \hat{\sigma}_n^-
    \label{eq:eom_H_wg}
\end{equation}

Meanwhile, the dissipation is described by the Linbladian $\mathcal{L}[\hat{\rho}]$:

\begin{equation}
\begin{split}
    \mathcal{L}[\hat{\rho}] &= \sum_{n,m} \Bigg ( -\frac{\Gamma_{nm}}{2} \hat{\sigma}_m^+ \hat{\sigma}_n^-  \hat{\rho} - \frac{\Gamma_{mn}}{2} \hat{\rho} \hat{\sigma}_m^+ \hat{\sigma}_n^-\\
    &+ \frac{\Gamma_{nm} + \Gamma_{mn}}{2} \hat{\sigma}_n^- \hat{\rho} \hat{\sigma}_m^+ \Bigg )
\end{split}
\label{eq:eom_L}
\end{equation}
When $n = m$, the Linbladian refers to the spontaneous decay, and when $n \neq m$, it describes the correlated decay due to the interaction through the waveguide. Equations (\ref{eq:eom_H_wg}) and (\ref{eq:eom_L}) together describe the waveguide quantum electrodynamics of the system. By\cite{PhysRevA.95.033818,PhysRevResearch.2.043213,PhysRevResearch.3.033233,PhysRevA.88.043806}, the exchange interaction strength $\Omega_{nm}$ and the coefficient $\Gamma_{nm}/2$ in the Linbladian are the real part and imaginary part of the Green's function matrix element from the transmon-transmon interaction through the waveguide. In a bidirectional 1D waveguide, the matrix element reads:
\begin{equation}
    \Omega_{nm} - i\frac{\Gamma_{nm}}{2} = -i\frac{\Gamma}{2} e^{ik\abs{z_n - z_m}},
\end{equation}
where $\abs{z_n - z_m}$ denotes the separation between transmon $n$ and $m$, $\Gamma$ is the decay rate of the individual transmons while $k$ is the wave number of the waveguide mode. 

By \cite{PhysRev.93.99,PhysRevA.98.063815}, if all the qubits are indistinguishable, the permutation symmetry determines the symmetry of the wavefunction and, in turn, facilitates the reduction of the state space. In our case, permutation symmetry applies when all the transmons are homogeneously coupled to the driving field and equally interact with each other. In the Hamiltonian, we can give all the transmons the same Rabi frequency and detuning:

\begin{equation}
    \Omega_n = \Omega, \quad \Delta_n = \Delta.
\end{equation}
In the Linbladian, we can let all the transmons be separated by an integer number of wavelengths, where $\Omega_{nm} = 0$ and $\Gamma_{nm} = \Gamma$:
\begin{equation}
    \mathcal{L}[\hat{\rho}] = \sum_{n, m}\frac{\Gamma}{2}(-\hat{\sigma}_m^+ \hat{\sigma}_n^- \hat{\rho} -\hat{\rho} \hat{\sigma}_m^+ \hat{\sigma}_n^- + 2\hat{\sigma}_n^- \hat{\rho} \hat{\sigma}_m^+).
\end{equation}
One can check that the interchange of indices $m$ and $n$ gives the same dynamics. When initialized in the ground state, the system will stay in the fully symmetric subspace with total spin $S = N/2$. Since every excited qubit in this subspace has the same phase, their photon emission can constructively interfere with each other, leading to superradiance. We say this system obeys the perfect (or pure) driven Dicke model. Since the transmons are indistinguishable in this case, we can introduce the collective spin operator:

\begin{equation}
    \hat{S}_+ = \sum_{n} \hat{\sigma}_n^+, \quad \hat{S}_- = \sum_{n} \hat{\sigma}_n^-, \quad \hat{S}_z = \frac{1}{2} \sum_n \hat{\sigma}_n^z
\end{equation}
so that the Hamiltonian and Linbladian in Eq. (\ref{eq:eom}) become:
\begin{equation}
    H = \frac{\Omega}{2}(\hat{S}_+ + \hat{S}_-) - \Delta \hat{S}_z 
    \label{eq:H_DDM}
\end{equation}
\begin{equation}
    \mathcal{L}[\hat{\rho}] = \frac{\Gamma}{2}(2\hat{S}_- \hat{\rho} \hat{S}_+ - \hat{S}_+ \hat{S}_- \hat{\rho} - \hat{\rho} \hat{S}_+ \hat{S}_-),
    \label{eq:L_DDM}
\end{equation}
where $H_{wg} = 0$. Because Eqs. (\ref{eq:H_DDM}) and (\ref{eq:L_DDM}) only involve total spin operators, they conserve the total spin of the system. In other words, density operators with the same total spin share the same dynamics and will reach the same steady state. This is elaborated in Sec. \ref{sec:explanation} and Appendix \ref{app:degenracy}.

\subsection{Perturbing the driven Dicke model\label{sec:pert_DDM}}
Given the dynamics of the perfect driven Dicke model, we want to investigate the perturbed dynamics of the system and compare with the unperturbed one. In general, there are three ways to perturb the driven Dicke model. One way is to introduce a dissipative (or $\Delta \mathcal{L}$ type) perturbation:
\begin{equation}
    \Delta \Dot{\hat{\rho}} = \Delta\mathcal{L}
    \label{eq:delta_rho_L}
\end{equation}
This type of perturbation leads to purely real contribution to the Louivillian superoperator which, in lowest order, contributes to the decay behavior. In this work, we investigate one example of this type of perturbation, the local dephasing with homogeneous dephasing rate for each transmon \cite{PhysRevA.98.063815}:
\begin{equation}
    \Delta\mathcal{L}[\hat{\rho}] = \sum_n \frac{\Gamma_\phi}{4}(\hat{\sigma}_n^z \hat{\rho} \hat{\sigma}_n^z - \hat{\rho}),
    \label{eq:Delta_L_phi}
\end{equation}
where $\Gamma_\phi$ is the dephasing rate. We note that this perturbation, only with the same dephasing rate for each transmon, preserves the permutation symmetry of the transmons.

Another type of perturbation is in the Hamiltonian:
\begin{equation}
    \Delta \Dot{\hat{\rho}} = -i [\Delta H, \hat{\rho}]
    \label{eq:delta_rho_H}
\end{equation}
This type of perturbation leads to purely imaginary contribution to the Louivillian superoperator which, in lowest order, contributes to oscillation and contributes to decay rate in second order. We investigate two examples of this type of perturbation: introducing individual driving phases
\begin{equation}
    \Delta H = \sum_n \frac{\Omega}{2} \left [ (e^{i\phi_n} - 1) \hat{\sigma}_n^+ + (e^{-i\phi_n} - 1) \hat{\sigma}_n^- \right ],
    \label{eq:pert_drv_phase}
\end{equation}
and introducing individual detunings
\begin{equation}
    \Delta H = \sum_n \frac{\Delta_n - \Delta}{2} \hat{\sigma}_n^z
\end{equation}

Lastly, we will discuss the perturbation from changing the separation of transmons, which is a combination of the $\Delta \mathcal{L}$ type and $\Delta H$ type perturbation:

\begin{equation}
\begin{split}
    \Delta H = \sum_{n, m} \Delta \Omega_{nm} \hat{\sigma}_m^+ \hat{\sigma}_n^-
\end{split}
\label{eq:delta_dot_rho_sep_delta_H}
\end{equation}

\begin{equation}
\begin{split}
    \Delta \mathcal{L}[\hat{\rho}] &= \sum_{n,m} \Bigg ( -\frac{\Delta \Gamma_{nm}}{2} \hat{\sigma}_m^+ \hat{\sigma}_n^-  \hat{\rho} - \frac{\Delta \Gamma_{mn}}{2} \hat{\rho} \hat{\sigma}_m^+ \hat{\sigma}_n^-\\
    &+ \frac{\Delta \Gamma_{nm} + \Delta \Gamma_{mn}}{2} \hat{\sigma}_n^- \hat{\rho} \hat{\sigma}_m^+ \Bigg )
\end{split}
\label{eq:delta_dot_rho_sep_delta_L}
\end{equation}
where $\Delta \Omega_{nm}$ and $\Delta \Gamma_{nm}/2$ are the real part and imaginary part of the perturbation in the Green's function matrix element due to the perturbation in the separation, which reads:
\begin{equation}
    \Delta \Omega_{nm} -i\frac{\Delta \Gamma_{nm}}{2} = -i\frac{\Gamma}{2}(e^{ik\Delta \abs{z_n - z_m}} - 1) .
    \label{eq:Delta_g_nm}
\end{equation}
For an equally spaced 1D lattice of transmons, with lattice constant perturbed by $\Delta d$, the Taylor expansion of Eq. (\ref{eq:Delta_g_nm}) to the second order shows that $\Delta \Omega_{nm} \propto \Delta d$ and $\Delta \Gamma_{nm} \propto \Delta d^2$.

\section{\label{sec:results}Results}
In this section, to show the effect of different perturbations, we first illustrate the driven Dicke superradiance in a 1D lattice, using homogeneous Rabi frequencies $\Omega_n = \Omega$, detunings $\Delta_n = \Delta = 0$, and equal separation between adjacent transmons $d = 1\lambda$ in Sec. \ref{sec:results_DDM}. In each later subsection, we keep everything the same except for one type of perturbation to compare the result with the unperturbed system. To guarantee the system is in the superradiant phase, according to \cite{ferioli2023non,PhysRevLett.127.243602}, we keep the ratio between the drive and the collective decay to be a constant ($\beta = 2 \Omega / (N \Gamma) = 4$, that is, $\Omega = 2N\Gamma$) for all the plots. We simulate for at most $8$ transmons and take the case for $N = 4$ as an example in several figures. For all four types of perturbations, superradiance is nearly unchanged at early times, but it gradually transitions to a distinct value asymptotically. The duration of the superradiant behavior increases as the size of the perturbation decreases, with the scaling of this duration depending on the type of perturbation. It is worth mentioning that the phenomena observed in the $N=4$ case also extend to systems with more transmons, up to the maximum number of transmons in our calculation, $N=8$. 

For all the plots of time evolution, unless otherwise stated, we scale the time $t$ to the decay rate $\Gamma$ and define the scaled time $\tau$ as:

\begin{equation}
    \tau = \Gamma t.
\end{equation}

For the observation, by \cite{gross1982superradiance,allen1987optical}, the photon emission rate consists of two parts:

\begin{equation}
    \gamma(t) = \gamma_0 (t) + \gamma_{SR}(t)
    \label{eq:gamma_tot}
\end{equation}
where the first term
\begin{equation}
    \gamma_0 (t) = \sum_n \Gamma \langle \hat{e}_n \rangle (t)
    \label{eq:gamma_0}
\end{equation}
refers to the individual spontaneous emission and the second term
\begin{equation}
    \gamma_{SR}(t) = \sum_{n \neq m} \Gamma_{nm} \langle \hat{\sigma}_m^+ \hat{\sigma}_n^- \rangle (t)
    \label{eq:gamma_SR}
\end{equation}
corresponds to the correlated photon emission. In this work, we focus on the correlated photon emission. When $\gamma_{SR} > 0$, the total photon emission rate is greater than the individual spontaneous photon emission rate and we consider the system to be superradiant, while $\gamma_{SR} < 0$ indicates the subradiance of the system. The case $\gamma_{SR} = 0$, where the system has only individual spontaneous emission, is considered as a boundary between superradiance and subradiance.

\subsection{Perfect driven Dicke model: superradiance in a 1D array\label{sec:results_DDM}}
\begin{figure}[!htbp]
  \centering
  \includegraphics[width = 0.5\textwidth]{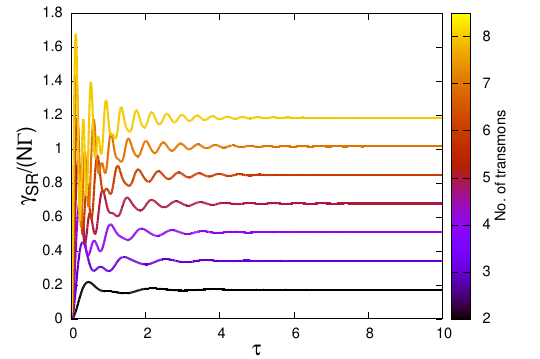}
  \caption{Time evolution of the normalized correlated photon emission rates $\gamma_{SR}/(N\Gamma)$ in Eq. (\ref{eq:gamma_SR}), with different qubit numbers, $N = 2-8$. The scaled time, $\tau = \Gamma t$.}
  \label{fig:ddmGamma_SR}
\end{figure}

To illustrate the driven Dicke superradiance, we calculate the time evolution of an $N$-qubit system, where $N$ ranges from $2$ to $8$ using the full density matrix equations. We start every transmon in the ground state. The time evolution of the correlated photon emission rate is shown in Fig. \ref{fig:ddmGamma_SR}.

Here we normalize the correlated photon emission rate by the decay rate, $\Gamma$, and the number of transmons, $N$. All the curves have fast oscillation at early times from the collective Rabi oscillation, and reach steady states after about $t = 10/\Gamma$. Note that the steady-state normalized correlated photon emission rate $\gamma_{SR}/(N\Gamma)$ is proportional to the qubit number, $N$. Such linear relation implies the quadratic relation between the total correlated photon emission rate $\gamma_{SR}$ and $N$, which is a signature of the driven Dicke superradiance. For the simulation with greater $N$, ranging from $5$ to $30$, this relation still holds. 

\subsection{Perturbation: local dephasing\label{sec:results_dephase}}

\begin{figure}[!htbp]
  \centering
  \includegraphics[width = 0.5\textwidth]{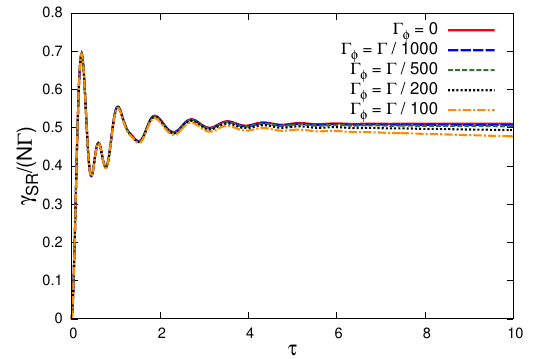}
  \caption{Early-time behavior of $\gamma_{SR}/(N\Gamma)$ with dephasing rates $\Gamma_{\phi} = 0$, $\Gamma/1000$, $\Gamma/500$, $\Gamma/200$, and $\Gamma/100$. The scaled time, $\tau = \Gamma t$}
  \label{fig:pop_dephase_early}
\end{figure}

To investigate the effect of small local dephasing, we compare the time evolution of the perturbed system with the perfect driven Dicke model for both early-time and long-term behavior. The local dephasing does not conserve total spin $S$, leading to evolution out of the $S = N/2$ subspace. For a 1D-lattice consisting of $4$ transmons, the early-time behavior of the correlated photon emission rates is shown in Fig. \ref{fig:pop_dephase_early} for different dephasing rates in Eq. (\ref{eq:Delta_L_phi}). 

At early times, the curves with different dephasing rates overlap because the chosen dephasing rates are much smaller than $\Gamma$ and $\Omega$. They gradually split from each other later. This strength of dephasing can be considered as a perturbation at early times because the size of the change in the signal is proportional to the strength of the perturbation. Similarly, the perturbations discussed below (driving phase, separation, and detuning) hardly affect the early-time superradiant behavior of the correlated photon emission rate when their strength is small. 

\begin{figure}[!htbp]
  \centering
  \includegraphics[width = 0.5\textwidth]{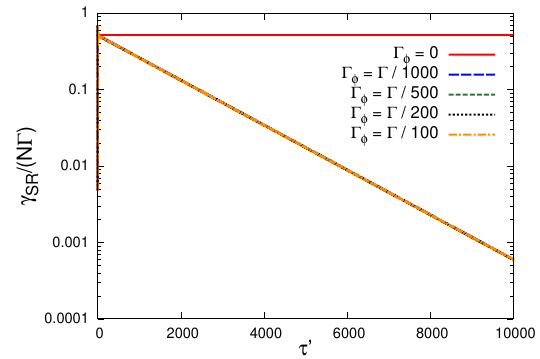}
  \caption{Evolution of $\gamma_{SR}/(N\Gamma)$ with different dephasing rates. We define $\tau^\prime = 1000 \Gamma_\phi t$, except for $\Gamma_\phi=0$ where $\tau^\prime = \Gamma t$.}
  \label{fig:pop_dephase_scaled}
\end{figure}

For the long-term behavior of the system, we plot the time evolution of each transmon's normalized correlated photon emission rate under different dephasing rates $\Gamma_\phi$ in Fig. \ref{fig:pop_dephase_scaled}.
We use the log scale for the y-axis. In addition, for each curve we scale the time $t$ by the dephasing rate $\Gamma_\phi$, to obtain the scaled time $\tau^\prime$ in Fig. \ref{fig:pop_dephase_scaled}.
The straight lines in the plot imply that the dephasing causes exponential decay of the correlated photon emission rate for this time scale. In other words, while the positive correlated photon emission rate can persist for a limited time, even a small dephasing rate ($\Gamma_\phi = 0.001\Gamma$) will eventually undermine the superradiance. With greater perturbation, the correlated photon emission rate decays faster. By scaling the time linearly to the dephasing rate, the curves overlap with each other. This implies that the magnitude of the perturbation does not significantly change the steady state, instead, it modulates the rate to reach the steady state. 

For more transmons, the phenomenon above also holds: the correlated photon emission rates exhibit exponential decay and, with the same scaled time as Fig. \ref{fig:pop_dephase_scaled}, the curves with different perturbation strengths collapse, which means the decay rate of $\gamma_{SR} / (N \Gamma)$ is proportional to the perturbation strength. We also fixed the dephasing rate $\Gamma_\phi = 0.01 \Gamma$, calculated the time evolution of $\gamma_{SR} / (N \Gamma)$ for $N = 3 \sim 8$, and evaluated the population living in the $S = N / 2$ subspace. The starting point of each curve's exponential decay is proportional to $N$, but the slope does not significantly depend on $N$ and the asymptotic value of $\gamma_{SR}/(N\Gamma)$ is approximately zero, regardless of $N$. The population in the $S = N / 2$ subspace decreases with $N$, which means local dephasing opens a path going out of the $S = N / 2$ subspace and this effect becomes more significant when $N$ increases.

\subsection{Perturbation: driving phase\label{sec:results_drv_phase}}

\begin{figure}[!htbp]
  \centering
  \includegraphics[width = 0.5\textwidth]{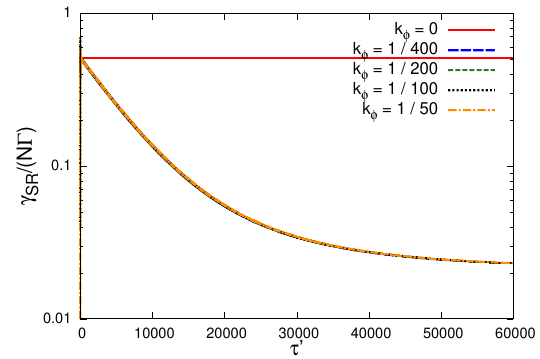}
  \caption{Same as Fig. \ref{fig:pop_dephase_scaled} except with different slopes of the driving phase. Except for the unperturbed ($k_\phi = 0$) case, where $\tau^\prime = \Gamma t$, we define $\tau^\prime = (400 k_\phi)^2 \Gamma t$.}
  \label{fig:pop_drv_phase_sclaed}
\end{figure}

Introducing a driving phase that depends on the transmon renders the system driven out of the $S = N / 2$ subspace. Here we investigate the effect of a linear driving phase perturbation:

\begin{equation}
    \phi_n = 2 \pi k_\phi n,
\end{equation}
where $n$ is the index of the transmon, $\phi_n$ is the phase introduced to the $n$th transmon, $k_\phi$ is the slope of the driving phase and is considered a measure of the perturbation. The time evolution of the correlated photon emission rate is shown in Fig. \ref{fig:pop_drv_phase_sclaed}. For each curve, we scale the original scaled time $\Gamma t$ by the square of the perturbation, as shown in the caption of the figure.

The result shows that even a tiny perturbation in driving phases changes the steady-state superradiance considerably, and a larger perturbation results in faster decay. With quadratically scaled time, the curves are on top of each other. This suggests that the perturbation in $k_\phi$ does not affect the steady state itself. Instead, it modulates the speed to reach the steady state quadratically. 

For the same calculation as Fig. \ref{fig:pop_drv_phase_sclaed} but with more transmons, the curves with smaller perturbation strengths overlap very well, but gradually deviate from each other when the perturbation strength grows. This effect becomes more significant when the number of transmon increases. One reason lies in the linear driving phase we introduce: this phase deviation accumulates with the number of transmons and will become too large to be considered as a perturbation when $N$ increases. When the perturbation strength is fixed at $k_\phi = 0.02$ but $N$ grows from $3$ to $8$, the starting point of $\gamma_{SR}/(N\Gamma)$ rises up, as implied by Fig. \ref{fig:ddmGamma_SR}, but the initial slope of the decay increases with $N$ and the asymptotic value scales approximately as $\sim 1/N$. Meanwhile, the population in the $S = N / 2$ declines with $N$, that is, the driving phase perturbation drives the system out of the $S = N / 2$ subspace and this is increasingly significant with larger $N$.

\subsection{Perturbation: separation\label{sec:results_sep}}
An imperfect separation also opens a path to drive the system out of the $S = N / 2$ subspace. To investigate the effect of the perturbation due to the separation, we compare the long-term behavior of the correlated photon emission rate of the perturbed system with the perfect one. For this perturbation, the transmons are equally separated with lattice constant $d$ but the separation is slightly different from a wavelength. The perfect driven Dicke case has $d = 1 \lambda$ while the perturbed lattice constants will be $d = 1 \lambda + \Delta d$, where $\Delta d$, the deviation from the integer multiple of the wavelength, is a measure of the perturbation. Note that the perturbation is a combination of $\Delta \mathcal{L}$ and $\Delta H$ type perturbation, as discussed in Sec. \ref{sec:pert_DDM}. By the Taylor expansion of Eq. (\ref{eq:Delta_g_nm}) to second order, $\Delta \Omega_{nm}$, related to the $\Delta H$ type perturbation, is linear in $\Delta d$; while $\Delta \Gamma_{nm}$, associated with the $\Delta \mathcal{L}$ type perturbation, is quadratic in $\Delta d$.

\begin{figure}[!hbp]
  \centering
  \includegraphics[width = 0.5\textwidth]{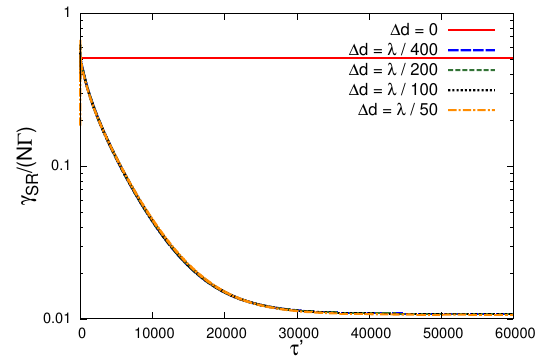}
  \caption{Same as Fig. \ref{fig:pop_dephase_scaled} except with different lattice constants. Except for the unperturbed ($\Delta d = 0$) case, where $\tau^\prime = \Gamma t$, we define $\tau^\prime = (400 \Delta d / \lambda)^2 \Gamma t$.}
  \label{fig:pop_sep_scaled}
\end{figure}

Figure \ref{fig:pop_sep_scaled} shows the time evolution of the correlated photon emission rate with perturbation in lattice constant $\Delta d = 0, \lambda / 400, \lambda / 200, \lambda / 100, \lambda / 50$. For each curve, we further scale its scaled time $\Gamma t$ by a factor proportional to the square of the perturbation, as shown in the caption of the figure.

Again, the perturbation in the lattice constant deteriorates the Dicke superradiance significantly and the greater perturbation leads to faster decay. In addition, scaling the time quadratically according to the perturbation brings all the curves to overlap. Like the previous two examples, the magnitude of the perturbation seems to not significantly change the steady-state value of the $\gamma_{SR}/(N\Gamma)$ but modulates the rate to reach the steady state quadratically. 

Similar to the driving phase case, for the same calculation as Fig. \ref{fig:pop_sep_scaled} except for more transmons, the curves corresponding to minor perturbation collapse but those with stronger perturbation gradually deviate. This effect becomes increasingly significant with larger $N$ because the maximum change in separation accumulates with the number of transmons and will become too large to be considered as a perturbation. When the perturbation strength is fixed at $\Delta d  = 0.02 \lambda$ and $N$ varies from $3$ to $8$, $\gamma_{SR}/(N\Gamma)$ starts declining at a value proportional to $N$ but with a steeper slope and settles at a lower value, which scales as approximately $\sim 1/N$. At the same time, the population in the $S = N / 2$ also goes down with higher $N$. This means the perturbation in separation couples the system outside of the $S = N / 2$ subspace and this effect is enhanced with more transmons. 

\subsection{An exceptional case: symmetric, and equally spaced detuning\label{sec:results_det}}
\begin{figure}[!htbp]
  \centering
  \includegraphics[width = 0.5\textwidth]{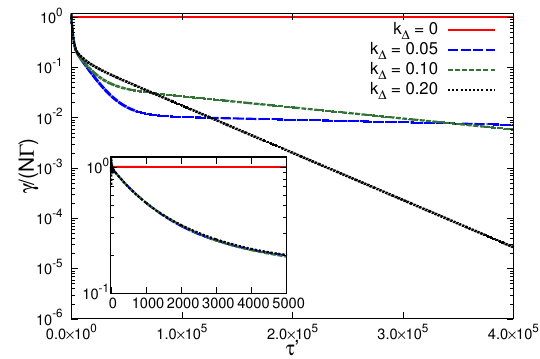}
  \caption{Similar to Fig. \ref{fig:pop_dephase_scaled} except with different slopes of the detuning and showing the scaled {\it total} photon emission rate. The inset shows the early-time behavior of $\gamma / (N\Gamma)$. Except for the unperturbed ($k_\Delta = 0$) case, where $\tau^\prime = \Gamma t$, we define $\tau^\prime = (k_\Delta / 0.05)^2 \Gamma t$.}
  \label{fig:pop_det_scaled}
\end{figure}

Inhomogeneous detuning also leads to coupling outside of the $S = N/2$ subspace. When the detuning has symmetric changes, the perturbation is shown to be an exceptional case due to the special configuration of detuning profiles. Here we will show the effect of linear detuning:
\begin{equation}
    \Delta_n = k_\Delta (n - N/2 + 1/2) \Gamma,
\end{equation}
where $n = 0, 1, 2,...N - 1$. The slope $k_\Delta$ is considered as a measure of the perturbation. In contrast to the above $3$ perturbations, the long-term behavior of the correlated photon emission rate is negative, implying that the system becomes subradiant. To explicitly show the evolution of the extent of subradiance, we plot the total photon emission rate $\gamma / (N\Gamma)$ instead of the correlated photon emission rate in Fig. \ref{fig:pop_det_scaled}. The system is subradiant when this scaled total photon emission rate is less than $\simeq 1/2$. For each curve, the original scaled time $\Gamma t$ is further scaled by a factor to the square of the corresponding slope $k_{\Delta}$.

Each curve shows a relative faster decay at early times, followed by an extremely slow tail. For the early-time behavior, each curve, with time scaled quadratically to the perturbation measured by the slope $k_\Delta$, overlaps with each other but splits later. For the asymptotic behavior, the total photon emission rate shows exponential decay. Note that even if the time is scaled quadratically to $k_\Delta$, the late-time exponential decay still grows with increasing $k_\Delta$. So this decay rate depends on the perturbation strength beyond the second order. Despite the extremely slow decay, we claim that the system becomes totally subradiant at steady state - the total photon emission rate $\gamma$ is approximately $0$. This can be verified by evaluating the steady-state density operator using the null eigenvectors of the Liouvillian matrix and Eq. (\ref{eq:vec_rho_t}) discussed in Sec. \ref{sec:diagonalization_Liouvillian}. The zero total photon emission rate has been explained by the theory in \cite{stannigel2012driven,ramos2014quantum,PhysRevA.91.042116} - because we introduce a symmetric profile of detunings so that for any transmon with detuning $\Delta_n$, there will be a corresponding transmon $n^\prime$ with the detuning $\Delta_{n^\prime} = -\Delta_n$. These two transmons can pair up to compose a so-called Dimer\cite{stannigel2012driven,ramos2014quantum,PhysRevA.91.042116}, which exhibits no coupling to the light field when they reach the steady state. Also, the experiment in \cite{lei2023many} reports the so-called collectively induced transparency(CIT), where for enough many inhomogeneous emitters in a cavity driven by an external field, the contribution of each emitter with a certain detuning $\Delta_n$ to the cavity field scales as $\propto \Delta_n^{-1}$ and can pair with the emitter with the same amplitude but opposite detuning so as to cancel each other. As a result, the negative correlated photon emission rate cancels the spontaneous emission rate so that the total photon emission in Eq. (\ref{eq:gamma_tot}) is zero. As for the steady-state correlated photon emission rates, we claim that they are different, unlike the perturbations in the previous sections. The reason lies in different detuning profiles induced by different $k_\Delta$, rendering different average excitations. By Eq. (\ref{eq:gamma_0}), the average spontaneous photon emission rates $\gamma_0/(N \Gamma)$ are equal to the average excitation and therefore are different. Because the total photon emission, which is the sum of the spontaneous photon emission rate and correlated photon emission rate, is zero, different spontaneous photon emission rate implies the different correlated photon emission rate.

With greater $N$, the decreasing of $\gamma_{SR}/(N\Gamma)$ starts at a higher value but with a sharper slope. The asymptotic behavior depends on whether $N$ is even or odd: when the transmon number is even, the tail becomes increasingly slow with larger $N$ and eventually becomes fully subradiant. In contrast, if the transmon number is odd, the system is not fully subradiant - the asymptotic total photon emission rate is not zero and decreases with $N$. The reason lies in the dangling transmon - the rest $N - 1$ transmons pair up to generate dimers and are effectively decoupled from the field, leaving one transmon that can still emit photons. When $N$ increases, due to our specific perturbation profile, the dangling transmon is increasingly detuned from the driving, rendering less excitation and photon emission. We also fixed $k_\Delta = 0.2$ and estimated the steady-state population in the $S = N / 2$ subspace for $N = 2 \sim 6$. In contrast to the previous $3$ kinds of perturbations, the population is at least one order of magnitude smaller and shows alternative behavior - the population for the even $N$ case is not only lower but also decreases faster than the odd $N$ case. This coincides with the alternative asymptotic behavior of the total photon emission rate. However, whether $N$ is even or not, the population declines, which means the symmetric, linear detuning perturbation paves the way to leak out of the $S = N / 2$ subspace and the effect is increasingly significant with more transmons.

\section{Explanation, an angular momentum coupling scheme\label{sec:explanation}}

In this section, we will give an explanation for the behavior noted in Sec. \ref{sec:results}. We first diagonalize the Liouvillian matrix and capture the features of the eigenvalues, revealing a degenerate subspace of steady states. Then we investigate the degenerate subspace using an angular momentum basis.

\subsection{Diagonalization of the Liouvillian\label{sec:diagonalization_Liouvillian}}
Considering that the Linbladian in the master equation is a superoperator, it is convenient to investigate the dynamics of the system if the equation of motion can be mathematically represented by a matrix-vector product form. This is particularly useful for time-independent Lindbladians like those studied here.

In addition to the master equation in Eq. (\ref{eq:eom}), a general form of the dynamics can be described by the Liouvillian:

\begin{equation}
    \frac{d\hat{\rho}}{dt} = \mathscr{L}[\hat{\rho}],
    \label{eq:eom_Liouvillian}
\end{equation}
where $\mathscr{L}[\cdot]$ is the Liouvillian superoperator. Thanks to the mathematical tool called Fock-Liouville
space\cite{thingna2021degenerated,manzano2020short}, we can reshape the density matrix $\hat{\rho}$ into a vector $\vec{\rho}$ and construct the $2^{2N} \times 2^{2N}$ Liouvillian matrix $\hat{\mathscr{L}}$ according to Eq. (\ref{eq:eom}) so that Eq. (\ref{eq:eom_Liouvillian}) becomes:

\begin{equation}
    \frac{d\vec{\rho}}{dt} = \hat{\mathscr{L}}\vec{\rho},
    \label{eq:eom_Liouvillian_vec}
\end{equation}

In general, the Liouvillian matrix is not hermitian and has complex eigenvalues associated to left eigenvectors and right eigenvectors satisfying biorthogonality. For the following discussion, we refer to right eigenvectors by ``eigenvectors'' unless otherwise stated. For a constant matrix $\hat{\mathscr{L}}$ (the cases we treat), the eigenvalues $\mu_n$ and the corresponding eigenvectors $\vec{\rho}_n$ imply the dynamics of the system:

\begin{equation}
    \vec{\rho}(t) = \sum_n c_n e^{\mu_n t} \vec{\rho}_n,
    \label{eq:vec_rho_t}
\end{equation}
where $c_n$ is the projection of the initial vector $\vec{\rho}(0)$ on the $n$th eigenvector $\vec{\rho}_n$. For each eigenvalue $\mu_n = \kappa_n + i\nu_n$ with $\kappa_n,\nu_n$ real, the non-positive real part indicates the decay while the imaginary part accounts for the oscillation behavior. Here we only plot the real parts that control the decay rate, which is closely related to the long-term behavior of the system. 

For a constant Liouvillian matrix $\hat{\mathscr{L}}$, there must be at least one eigenvalue $\mu_n = 0$ with the corresponding null eigenvector. By Eq. (\ref{eq:vec_rho_t}), when $t \rightarrow \infty$, all the eigenvectors with negative real parts decay to zero. So the linear combination of the null eigenvectors $\vec{\rho}_n$ yields the steady-state density operator. When there is more than one null eigenvector, the overlaps together with null eigenvectors determine the steady state. However, if there is only one null eigenvector, the steady state is solely determined by the null eigenvector, which is the only eigenvector with non-zero trace. Specifically, if the system is prepared so that its density operator can be reshaped into a null eigenvector, the system will stay in this state forever. The steady-state spontaneous photon emission rate and correlated photon emission rate can be evaluated by Eqs. (\ref{eq:gamma_0}) and (\ref{eq:gamma_SR}), which will indicate whether the steady state is superradiant or subradiant. There are a large number of eigenstates with $\mu_n=0$ for the perfect driven Dicke model indicating a large subspace of possible steady states. Thus, the steady state for the perfect driven Dicke model depends on the initial state which we have taken to be the state with all transmons in their ground state. It is the degenerate, steady state subspace that is the key to understanding the behavior noted in Sec. \ref{sec:results} because the steady state depends on degenerate perturbation theory within the superoperator. For the systems in Secs. III B,C,D, a nonzero perturbation leads to a single null eigenvector so that the steady state is independent of the initial conditions.

For perturbations of the driven Dicke model, it is the perturbation theory within the degenerate null subspace that determines the long-time behavior. As is discussed in Sec. \ref{sec:pert_DDM}, there are two types of perturbations: dissipative (or $\Delta \mathcal{L}$ type) perturbation and $\Delta H$ type perturbation, each contributing to purely real (by Eq. (\ref{eq:delta_rho_L})) and purely imaginary perturbations (by Eq. (\ref{eq:delta_rho_H})), respectively.  For purely real perturbations, the eigenvalues will have a real part proportional to the perturbation strength. For purely imaginary perturbations, the eigenvalues will have imaginary part proportional to the perturbation strength and real part quadratic in the strength because the real part arises from second order in the perturbation.

\begin{figure}[!htbp]
  \centering
  \includegraphics[width = 0.5\textwidth]{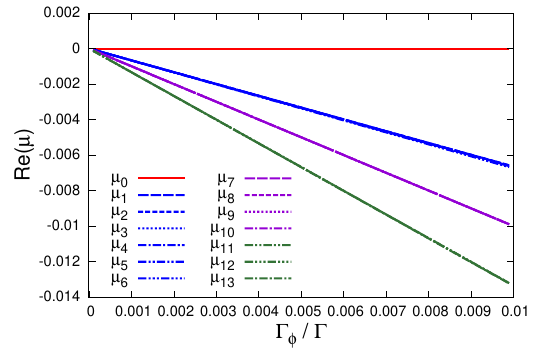}
  \caption{The cluster of 14 eigenvalues of the Liouvillian with the least negative real parts for the dephasing perturbation.}
  \label{fig:eig_dephase}
\end{figure}

\begin{figure}[!htbp]
  \centering
  \includegraphics[width = 0.5\textwidth]{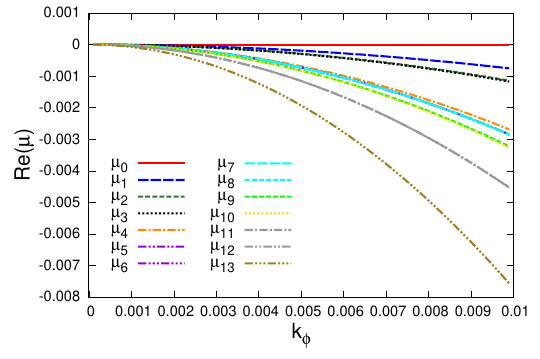}
  \caption{Same as Fig. \ref{fig:eig_dephase} except for the driving phase perturbation.}
  \label{fig:eig_drv_phase}
\end{figure}

\begin{figure}[!htbp]
  \centering
  \includegraphics[width = 0.5\textwidth]{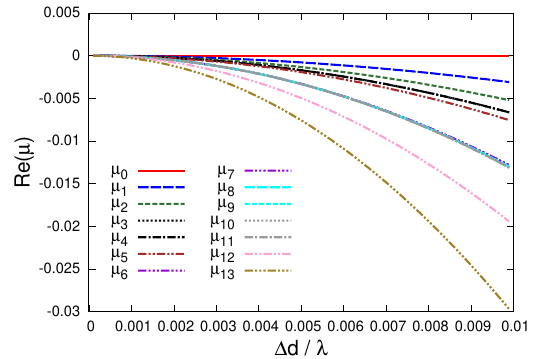}
  \caption{Same as Fig. \ref{fig:eig_dephase} except for the separation perturbation.}
  \label{fig:eig_sep}
\end{figure}

For a 4-transmon system, out of $(2^4)^2 = 256$ eigenvalues, we only plot the cluster of 14 eigenvalues with the smallest real parts. We chose to plot these because the perfect driven Dicke model with $N=4$ has $14$ eigenvalues with $\mu_n=0$. The other eigenvalues, with more negative real parts, decay quickly and are not closely related to the long-term behavior. The eigenvalues are illustrated in Figs. \ref{fig:eig_dephase}, \ref{fig:eig_drv_phase} and, \ref{fig:eig_sep} for the perturbation of dephasing, driving phase, and separation, respectively. 

For dephasing, the $14$ eigenvalues corresponding to the states represented in Fig. \ref{fig:eig_dephase} are pure real, with imaginary parts in the numerical diagonalization of the order of $\sim \!\! 10^{-15}$. In addition, the eigenvalues are clustered into four values (from the top to bottom) in Fig. \ref{fig:eig_dephase}, with the degeneracy of $1$, $6$, $4$, and $3$.
In Figs. \ref{fig:eig_drv_phase} and \ref{fig:eig_sep}, there is less degeneracy in the real part. For the driving phase perturbation, there are 3 pairs of eigenvalues with the same real parts: $Re(\mu_5) = Re(\mu_6) \neq Re(\mu_7) = Re(\mu_8) \neq Re(\mu_{11}) = Re(\mu_{12})$. However, each pair of these eigenvalues are mutually conjugate, with opposite sign imaginary parts. The other $8$ eigenvalues are different and pure real. For the separation perturbation, there are $4$ pairs of conjugate eigenvalues with different real parts: $\mu_3$ and $\mu_4$, $\mu_6$ and $\mu_7$, $\mu_8$ and $\mu_9$, $\mu_{10}$ and $\mu_{11}$. The other $6$ eigenvalues are pure real and nondegenerate. So the imaginary parts in the driving phase and separation cases eliminate the degeneracy in real parts. 

There are two main features in the plots. First, there is a qualitative difference between Fig. \ref{fig:eig_dephase} and Figs. \ref{fig:eig_drv_phase} and \ref{fig:eig_sep}. The real parts of eigenvalues are linear in the perturbation for the dephasing case. By Eq. (\ref{eq:vec_rho_t}), the negative real parts indicate the decay rate for the corresponding eigenvector and, in turn, the dynamics of the density operator. The linear dependence of the real parts on the perturbation strength means the decay rates contributing to the dynamics of the density operator are also linear to the perturbation strength. This accounts for the linear dependence of the decay rate of $\gamma_{SR}/(N\Gamma)$ on the dephasing rate $\Gamma_\phi$ in Fig. \ref{fig:pop_dephase_scaled}. Similarly, the quadratic dependence of real parts of eigenvalues on the perturbation for the driven phase and separation cases, Figs. \ref{fig:eig_drv_phase} and \ref{fig:eig_sep}, accounts for the quadratic relation between the decay rate of $\gamma_{SR}/(N\Gamma)$ and the perturbation in the two cases ($\Delta d$ and $k_\phi$). Second, there is only one eigenvalue $\mu_n = 0$ for non-zero perturbation and, when the perturbation approaches $0$, all the $14$ eigenvalues become degenerate $\mu_0 = \mu_1 = ... = \mu_{13} = 0$. Since a zero eigenvalue means the corresponding eigenvector does not evolve under the equation of motion in Eqs. (\ref{eq:eom_Liouvillian}) and (\ref{eq:vec_rho_t}), the degeneracy of the eigenvalues with $\mu_n = 0$, $N_{ss}$, indicates the possible number of steady-state density operators when there is no perturbation. To investigate the dependence of the degeneracy on the transmon number, we diagonalize the Liouvillian for $N$ ranging from $2$ to $7$ and Table \ref{tab:N_Degeneracy} illustrates the degeneracy $N_{ss}$ as a function of the transmon number $N$. 

\begin{table}[]
    \centering
    \caption{Relation between transmon number $N$ and the degeneracy of the eigenvalues of the Liouvillian $N_{ss}$}
    \begin{tabular}{c c}
    \hline
        $N$ & $N_{ss}$ \\
        \hline
        2 & 2\\
        3 & 5\\
        4 & 14\\
        5 & 42\\
        6 & 132\\
        7 & 429\\
    \hline
    \end{tabular}
    \label{tab:N_Degeneracy}
\end{table}

In the next section, we use an angular momentum coupling basis to count the dimension of the subspace $N_{ss}$ and evaluate the matrix elements of the perturbation-induced coupling matrix $\hat{C}$, in Eqs. (\ref{eq:C_eta_eta_prime_dephase}), (\ref{eq:vec_R_dot_C}), (\ref{eq:R_dot_C}), and (\ref{eq:C_eta_eta_prime_H}), within the subspace using the degenerate perturbation theory. Like usual degenerate perturbation theory, when the perturbation is small and not zero, the eigenvectors of $\hat{C}$ (steady states in the degenerate subspace) do not depend on the perturbation strength while the eigenvalues are proportional to the perturbation to the lowest order because the matrix $\hat{C}$ is proportional to the perturbation. For three of the perturbations considered, only one eigenvalue remains zero while the others have negative real parts, implying only one steady state with the others decaying away. This is fundamentally different from the perfect driven Dicke model, where the $N_{ss}$ eigenvalues are all zero, and explains why an infinitesimal perturbation qualitatively changes the steady state from that of the perfect driven Dicke model.

\begin{figure}[!htbp]
  \centering
  \includegraphics[width = 0.5\textwidth]{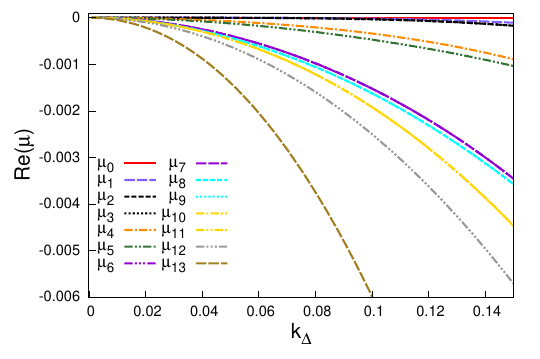}
  \caption{Same as Fig. \ref{fig:eig_dephase} except for the detuning perturbation. This range of $k_\Delta$ is beyond where lowest order perturbation is accurate to show the splitting of the four smallest eigenvalues.}
  \label{fig:eig_det}
\end{figure}

The symmetric and equally spaced detuning case is the exception. The steady-state density operator, indicated by the unique null eigenvector, is a pure state where the transmons form dimers. Because dimers do not couple to the waveguide mode, the total photon emission rate in Eq. (\ref{eq:gamma_tot}) is approximately $0$. In Fig. \ref{fig:eig_det}, despite the uniqueness of the null eigenvalue, there are $3$ pure real eigenvalues ($\mu_1$, $\mu_2$, and $\mu_3$) very close to it - they are almost degenerate to the null eigenvalue until around $k_\Delta > 0.13$. We find that these three eigenvalues are proportional to $k_\Delta^4$, indicating that the long-term decay in Fig. \ref{fig:pop_det_scaled} arises from fourth order in perturbation theory. This explains the extremely slow tails for $k_\Delta = 0.05$ and $k_\Delta = 0.1$ compared to the relatively fast decay for $k_\Delta = 0.2$ in Fig. \ref{fig:pop_det_scaled}. Different detuning profiles lead to different average excitations and account for the different steady-state correlated photon emission rates, as discussed below Fig. \ref{fig:pop_det_scaled}. Note that there are $4$ pairs of eigenvalues with the same real parts: $\mu_2$ and $\mu_3$, $\mu_6$ and $\mu_7$, $\mu_8$ and $\mu_9$, $\mu_{10}$ and $\mu_{11}$. Except for $\mu_2$ and $\mu_3$, which are real and degenerate, all the other $3$ pairs of eigenvalues are mutually conjugate.



\subsection{An angular momentum coupling scheme}
To investigate the dimension of the degenerate subspace, we start from the degeneracy of the dynamics of the perfect driven Dicke model represented by the toal spin operators in Eq. (\ref{eq:eom_DDM}). The angular momentum coupling basis is very useful when it comes to the dynamics described by total spin operators. For the dependence of the degeneracy on the transmon number, we use the eigenstates of the operators $\{\hat{S}_1^2, \hat{S}_{12}^2, \hat{S}_{13}^2, ..., \hat{S}_{1N}^2, \hat{S}_z\}$, where $\hat{S}_{1n}^2 = (\hat{S}_{1} + \hat{S}_{2} + ... + \hat{S}_n)^2$. Given $S_n = \frac{1}{2}$ and $\hat{S}_{1N} = \hat{S}$, the eigenstates can be written as:

\begin{equation}
   \ket{a, S, M} = \ket{(((\frac{1}{2}, \frac{1}{2}), S_{12}, \frac{1}{2}), S_{13}, \frac{1}{2}), ..., S, M}
\end{equation}
where $S$ and $M$ refer to the total spin and its projection on the $z$ axis. 
The index $a$ is used to distinguish between different angular momentum couplings with the same $S$ and $M$. In Table \ref{tab:Sa} in Appendix \ref{app:degenracy} we give an example of how the angular momentum couplings are labeled for $N=4$. The angular momentum couplings are orthogonal to each other, implying the definition of a basis of density operator using their outer product:
\begin{equation}
\begin{split}
    &Tr(\ket{a, S, M}\bra{a^\prime, S^\prime, M^\prime}) = \bra{a^\prime, S^\prime, M^\prime}\ket{a, S, M} \\
    &= \delta_{a a^\prime}\delta_{S S^\prime}\delta_{M M^\prime} = (\prod_{\nu = 2}^{N-1} \delta_{S_{1\nu} {S^{\prime}}\!\!_{1\nu}})\delta_{S S^\prime}\delta_{M M^\prime}
\end{split}
\label{eq:aSm_orthonormal}
\end{equation}

Given the total spin $S$ and the angular momentum coupling indices $a$ and $b$, the steady-state density operator for the unperturbed driven Dicke model, $\hat{\rho}_S^{a, b}$, can be written as:

\begin{equation}
    \hat{\rho}_S^{a, b} = \sum_{M, M^\prime} \rho^S_{M, M^\prime} \ket{a, S, M} \bra{b, S, M^\prime},
    \label{eq:rho_S_a_b}
\end{equation}
where the coefficients $\rho^S_{M, M^\prime}$ are not labeled with index $a$ and $b$ because of the degeneracy of the pure driven Dicke model (See Appendix \ref{app:degenracy}): The equations of motion in Eq. (\ref{eq:eom_DDM}) only change $M$ and conserve the total spin $S$. Therefore, the density operators with the same $S$ share the same equations of motion and must have the same coefficients $\rho^S_{M, M^\prime}$. The number of possible choices of $\hat{\rho}_S^{a, b}$ indicates the degeneracy of density operators given a total spin $S$. It is worth mentioning that the angular momentum coupling $a$ and $b$ are not necessarily the same and Eq. (\ref{eq:aSm_orthonormal}) implies the trace of $\hat{\rho}_S^{a, b}$:

\begin{equation}
    Tr(\hat{\rho}_S^{a, b}) = \delta_{a b}
\end{equation}
So, in addition to what we call ``diagonal'' operators with $a = b$ and trace $Tr(\hat{\rho}_S^{a, b}) = 1$, there is another type called ``off-diagonal'' operators with $a \neq b$ and trace being $0$. Due to the non-hermiticity and zero trace, "off-diagonal" operators themselves cannot be physical density operators, but together with the "diagonal" operators, they can constitute the steady-state density operator of the pure driven Dicke model. 

If $a$ or $b$ can take $D_S$ values, or the degeneracy for the state $\ket{S, m}$ is $D_S$, there should be $D_S^2$ possible steady-state density operators associated with the total spin $S$. By \cite{PhysRev.93.99,PhysRevA.96.023863,PhysRevA.98.063815}, the expression of $D_S$ for $N$ transmons with total spin $S$ reads:

\begin{equation}
    D_S = (2S + 1)\frac{N!}{(\frac{N}{2} + S + 1)!(\frac{N}{2} - S)!}
    \label{eq:D_S}
\end{equation}

So the total degeneracy (or the possible number of steady states) $N_{ss}$ is:
\begin{equation}
    N_{ss} = \sum_S D_S^2 = \frac{(2N)!}{N!(N + 1)!}
    \label{eq:N_ss}
\end{equation}

\begin{figure}[!htbp]
  \centering
  \includegraphics[width = 0.5\textwidth]{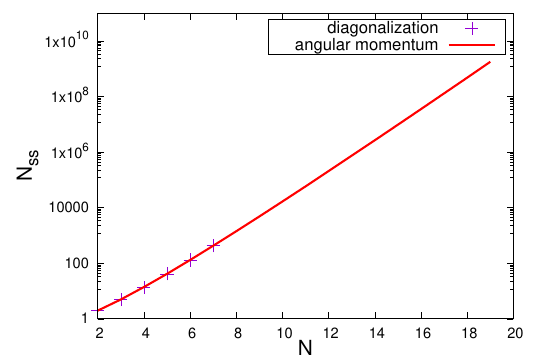}
  \caption{Comparison between the degeneracy from the diagonalization of the Liouvillian and the theoretical prediction by the angular momentum method, Eqs (\ref{eq:N_ss}) and (\ref{eq:D_S}).}
  \label{fig:N_ss}
\end{figure}

Due to the factorial in Eq. (\ref{eq:D_S}), the degeneracy increases rapidly with $N$. Using Eq. (\ref{eq:N_ss}) and (\ref{eq:D_S}), we calculate the degeneracy and compare it to the results from the diagonalization of the Liouvillian in Fig. \ref{fig:N_ss}. The numerical diagonalization of the Liouvillian shows perfect agreement to the theoretical prediction using the angular momentum basis and both grow somewhat faster than exponentially. This justifies our choice of the angular momentum coupling basis to represent the degenerate subspace. In the following discussion, we will use the angular momentum basis to describe the effect of perturbed dynamics in the degenerate subspace.

To simplify the notation, we attach an index $\eta$ to the combined indices $\{S, a, b\}$ so that density operators $\hat{\rho}_S^{a, b}$ can be written as $\hat{\rho}_\eta$. 

Given the degenerate subspace of steady states spanned by the $N_{ss}$ density operators $\{\hat{\rho}_\eta\}$, we investigate the effect of the perturbation on the subspace: With the definition of the left operator in Eq. (\ref{eq:rho_eta_L}) and the orthogonality relation in Eq. (\ref{eq:rho_eta_L_orth}), we project the superoperator $\mathscr{L}[\cdot]$ into the degenerate subspace, represented by an $N_{ss} \times N_{ss}$ matrix $\hat{C}$, whose matrix elements $C_{\eta \eta^\prime}$ denote the coupling from $\hat{\rho}_{\eta^\prime}$ to the time derivative $\dot{\hat{\rho}}_\eta$. The details of evaluating the matrix elements are elaborated in Appendix \ref{app:evaluation}. 

For the dephasing case, a pure dissipative perturbation, we diagonalize the matrix $\hat{C}$ and the comparison with the real parts of the eigenvalues from the diagonalization of the Liouvillian shows good agreement. So the dephasing perturbation causes a mixing between the degenerate steady states so that the eigenvalues, except for one with $\mu_0 = 0$, have negative real parts linearly dependent on the dephasing rate. As a result, there only remains one steady state and others decay away. The rate of decay can be retrieved from the real parts of the coupling matrix $\hat{C}$. As for the imaginary parts, they are all zero in the eigenvalues of the coupling matrix $\hat{C}$. So pure dissipative perturbation like dephasing introduces the first-order correction to the real parts of the eigenvalues. By Eq. (\ref{eq:Delta_L_phi}), the dephasing perturbation $\Delta \mathcal{L}$ is linearly proportional to the dephasing rate $\Gamma_\phi$. This accounts for the linear dependence of the decay rate of $\gamma_{SR}$ on the dephasing rate.

For the pure Hamiltonian-type perturbation, we diagonalize the matrix $\hat{C}$ for the driving phase case. This time all the real parts are zero while the imaginary parts agree well with the full diagonalization results. So the pure Hamiltonian-type perturbation introduces the first-order perturbation to the imaginary parts of the Liouvillian eigenvalues. It also leads to a second-order correction to the real parts. This accounts for the quadratic dependence of the decay rate of $\gamma_{SR}$ on the perturbation $k_\phi$.

The perturbation of separation, by Eqs. (\ref{eq:delta_dot_rho_sep_delta_H}), (\ref{eq:delta_dot_rho_sep_delta_L}), and (\ref{eq:Delta_g_nm}), is a combination of $\Delta H$ and $\Delta \mathcal{L}$ type perturbation. As mentioned above, $\Delta H$ type perturbation leads to a first-order change in the imaginary parts of the Liouvillian eigenvalues and a second-order correction in the real parts. Since the corresponding coefficients $\Omega_{nm} \propto \Delta d$, the contribution to the imaginary parts is proportional to $\Delta d$ and the contribution to the real parts is proportional to $\Delta d^2$. Meanwhile, the $\Delta \mathcal{L}$ type perturbation yields a first-order perturbation in the real parts and a second-order perturbation in the imaginary parts. But because the coefficients $\Gamma_{nm} \propto \Delta d^2$, the first-order perturbation in real parts is proportional to $\Delta d^2$ which is the same power of $\Delta d$ from the perturbation in imaginary parts. So both $\Delta \mathcal{L}$ type and $\Delta H$ type perturbations contribute to the perturbation in the real parts $\propto \Delta d^2$. This accounts for the quadratic dependence of the decay rate of $\gamma_{SR}$ on the perturbation $\Delta d$.




\section{\label{sec:summary}Summary}
We investigated the effect of perturbations on driven Dicke superradiance in a 1D transmon lattice coupled to a 1D wave\-guide. With perfect homogeneous driving and separation of exactly an integer multiple of the wavelength, the 1D lattice of transmons can exhibit driven Dicke superradiance. However, even an infinitesimal perturbation can lead to a significantly different steady state. 

To study the dynamics of the perturbed driven Dicke model, we diagonalized the Liouvillian for different sizes of perturbation, which reveals a degenerate subspace of steady states for the unperturbed system. Using an angular momentum coupling basis and first-order perturbation theory on the degenerate subspace, we obtained the expression of the degeneracy and construct the coupling matrix $\hat{C}$, the projection of the Liouvillian matrix $\hat{\mathscr{L}}$ in an angular momentum basis. In degenerate perturbation theory, the eigenstates (steady states in the degenerate subspace) do not depend on the perturbation strength but are mixed by the perturbation, leading to the modulation of the eigenvalues proportional to the perturbation to lowest order. The diagonalization of $\hat{C}$ shows the details of the modulation for the dephasing, driving phase, and separation perturbation. 

A pure dissipative perturbation, like dephasing, introduces a first-order correction to the real parts of the full Liouvillian's eigenvalues. This explains the linear dependence of the decay rate of $\gamma_{SR}$ on the dephasing rate. In contrast, a pure Hamiltonian type perturbation, like the driving phase, contributes a first-order correction to the imaginary parts and a second-order correction to the real parts. This accounts for the quadratic dependence of the decay rate of $\gamma_{SR}$ on the driving phase perturbation, $k_\phi$. The separation perturbation is a combination of pure dissipative and Hamiltonian type perturbation. Its Hamiltonian type perturbation, the strength of the perturbed exchange interaction, $\Delta \Omega_{nm}$, is linearly proportional to the change in lattice constant $\Delta d$. Its first-order correction contributes to the imaginary parts $\propto \Delta d$ while the second-order correction leads to real parts $\propto \Delta d^2$. The dissipative perturbation due to the change of separation, evaluated by $\Gamma_{nm}$, is quadratic to $\Delta d$. Its first-order correction contributes to the real parts $\propto \Delta d^2$. So both the Hamiltonian type perturbation and dissipative perturbation contribute to the real parts $\propto \Delta d^2$ - this accounts for the quadratic dependence of the decay rate of $\gamma_{SR}$ on $\Delta d$. For all three perturbations, there is one unique steady state and the others decay away. As with the more familiar degenerate perturbation theory for hermitian operators, the degenerate subspace of steady states explains why even an infinitesimal perturbation leads to complete mixing within the subspace. The exception was the detuning perturbation with equally spaced detunings. For this case, the perturbation does not fully lift the degeneracy of the steady states which led to a dependence of the steady state on the size of the perturbation.

For future research, several avenues remain open for exploration in the context of the driven Dicke model. One promising direction is the application of higher-order perturbation theory to better understand the dynamics of the imperfect driven Dicke model. Additionally, the study of novel phase transitions resulting from the interplay between parameterized driving and dissipation could reveal new and interesting phenomena \cite{PhysRevLett.114.040402,PhysRevA.89.023616,PhysRevA.97.053616,PhysRevLett.110.257204,PhysRevResearch.3.L022020,parmee2020signatures,PhysRevA.95.053833,PhysRevLett.107.113602,PhysRevA.96.053857,PhysRevLett.117.173601}. Another area worth investigating is the impact of additional factors on driven Dicke superradiance. For instance, incorporating exchange interactions into the model might provide valuable insights\cite{PhysRevLett.113.220502,PhysRevApplied.10.054062,niskanen2007quantum,hime2006solid}. Beyond continuous wave driving, exploring the driven Dicke model with quantized photons presents another intriguing research opportunity.  Finally, introducing disorder into the system could lead to novel physical behaviors and is a topic of active investigation\cite{PhysRevA.109.013715,PhysRevA.108.023708,PhysRevA.106.053703,lei2023many}.

\begin{acknowledgements}

This work was supported by the National Science Foundation under Award No. 2109987-PHY. This research was supported in part through computational resources provided by Information Technology at Purdue University, West Lafayette, Indiana.

\end{acknowledgements}

\appendix
\section{Degeneracy of the Liouvillian's eigenvalues, counting the number of possible steady states\label{app:degenracy}}
To discuss the degeneracy of Liouvillian's eigenvalues, we need to study the degeneracy of the dynamics of the perfect driven Dicke model\cite{ferioli2023non,PhysRevA.98.063815}:

\begin{equation}
\begin{split}
     \Dot{\hat{\rho}} &= -i [\frac{\Omega}{2}(\hat{S}_+ + \hat{S}_-) - \Delta \hat{S}_z, \hat{\rho}] \\
     &+ \frac{\Gamma}{2}(2\hat{S}_- \hat{\rho} \hat{S}_+ - \hat{S}_+ \hat{S}_- \hat{\rho} - \hat{\rho} \hat{S}_+ \hat{S}_-)
\end{split}
\label{eq:eom_DDM}
\end{equation}

Since all the operators in Eq. (\ref{eq:eom_DDM}) are total spin operators, they commute with the spin square operators. We consider the eigenstates of the operators $\{\hat{S}_1^2, \hat{S}_{12}^2, \hat{S}_{13}^2, ..., \hat{S}_{1N}^2, \hat{S}_z\}$:

\begin{equation}
    \ket{a, S, M} = \ket{(((\frac{1}{2}, \frac{1}{2}), S_{12}, \frac{1}{2}), S_{13}, \frac{1}{2}), ..., S, M},
    \label{eq:aSm}
\end{equation}
where the left-hand side is the in-short representation of the eigenstates using the index $a$ to distinguish between different angular momentum couplings. In Table \ref{tab:Sa}, we give an example of the state list for the transmon number $N = 4$.

\renewcommand{\arraystretch}{1.5}
\begin{table}[!htbp]
    \centering
    \caption{State table using index $a$ to distinguish between the degenerate states with the same total spin $S$}
    \begin{tabular}{c| c c c c |c}
    \hline
        $S$,$a$ & $S_1$ & $S_{12}$ & $S_{13}$ & S & M\\
        \hline
        $S = 0$, $a = 0$ & $\frac{1}{2}$ & 0 & $\frac{1}{2}$ & 0 & 0\\
        $S = 0$, $a = 1$ & $\frac{1}{2}$ & 1 & $\frac{1}{2}$ & 0 & 0\\
        \hline
        $S = 1$, $a = 0$ & $\frac{1}{2}$ & 0 & $\frac{1}{2}$ & 1 & -1, 0, 1\\
        $S = 1$, $a = 1$ & $\frac{1}{2}$ & 1 & $\frac{1}{2}$ & 1 & -1, 0, 1\\
        $S = 1$, $a = 2$ & $\frac{1}{2}$ & 1 & $\frac{3}{2}$ & 1 & -1, 0, 1\\
        \hline
        $S = 2$, $a = 0$ & $\frac{1}{2}$ & 1 & $\frac{3}{2}$ & 2 & -2, -1, 0, 1, 2\\
        \hline
    \end{tabular}
    \label{tab:Sa}
\end{table}

Note that the equation of motion only changes $M$ and conserves the total spin $S$, so the density operators with the same total spin $S$ share the same equations of motion. If we only consider the $S$ and $M$ degrees of freedom and denote the density operator with total spin $S$ by:
\begin{equation}
    \hat{\rho}^S = \sum_{M, M^\prime} \ket{S, M}\bra{S, M^\prime} \rho^S_{M, M^\prime},
\end{equation}
the dynamics of the matrix element $\rho^S_{M, M^\prime}$ can be written as:

\begin{equation}
\begin{split}
    &\Dot{\rho}^S_{M, M^\prime} = -i [\frac{\Omega}{2}(\!A_{M - 1} \rho^S_{M - 1, M^\prime} + A_{M} \rho^S_{M + 1, M^\prime}\\
    &- A_{M^\prime - 1} \rho^S_{M, M^\prime - 1} - A_{M^\prime} \rho^S_{M, M^\prime +1}\!) - \Delta (M \!-\! M^\prime)\rho^S_{M, M^\prime}]\\
    &+ \frac{\Gamma}{2}(2 A_M A_{M^\prime} \rho^S_{M + 1, M^\prime + 1} \!-\! A_{M - 1}^2 \rho^S_{M, M^\prime} \!-\! A_{M^\prime - 1}^2 \rho^S_{M, M^\prime}),
\end{split}
\label{eq:eom_rho^S}
\end{equation}
where
\begin{equation}
    A_M = \sqrt{S(S + 1) - M(M + 1)}
\end{equation}
Solving Eq. (\ref{eq:eom_rho^S}) yields the steady-state solution $\hat{\rho}^S_{ss}$. In this work, we are mainly interested in the steady-state
solution, so for the later discussion the matrix element of $\hat{\rho}_{ss}^S$ will be denoted by $\rho_{M, M^\prime}^S$. Taking the angular momentum coupling into account, for a given index $a$ for ket and $b$ for bra, the steady-state density operators can be written as:

\begin{equation}
    \hat{\rho}_S^{a, b} = \sum_{M, M^\prime} \rho^S_{M, M^\prime} \ket{a, S, M} \bra{b, S, M^\prime}.
    \label{eq:rho_eta}
\end{equation}
Because the equation of motion for the perfect driven Dicke model in Eq. (\ref{eq:eom_DDM}) and (\ref{eq:eom_rho^S}) does not depend on the angular momentum coupling, any choice of $a$ and $b$ will share the same coefficients $\rho^S_{M, M^\prime}$, which is the origin of the degeneracy of the possible steady-state density operators. Given total spin $S$ and transmon number $N$, the degeneracy of the angular momentum coupling degrees of freedom $D_S$ is shown in Eq. (\ref{eq:D_S}).



Because both $a$ and $b$ have $D_S$ degeneracy, the degeneracy of the density operator $\hat{\rho}_S^{a, b}$ should be $D_S^2$. Summing over $S$ gives the degeneracy of the eigenvalues of the Liouvillian in Eq. (\ref{eq:N_ss}). For example, for $N=4$, the degeneracy $N_{ss} = 2^2 + 3^2 + 1^2=14$ which matches the value in Table \ref{tab:N_Degeneracy}. 

\section{Perturbation theory in the degenerate subspace, evaluation of the coupling matrix elements\label{app:evaluation}}

Using the density operators in Eq. (\ref{eq:rho_S_a_b}) as a basis, the steady-state density operator for the perfect driven Dicke case can be written as a linear combination of the density operators in Eq. (\ref{eq:rho_S_a_b}):

\begin{equation}
    \hat{\rho}_{ss} = \sum_{S, a, b} R_{S, a, b} \hat{\rho}_S^{a, b} = \sum_{\eta} R_{\eta} \hat{\rho}_\eta
    \label{eq:rho_ss}
\end{equation}
where we use a single index $\eta$ to denote the 3 indices $\{S, a, b\}$ for simplicity. As an analogy to the left and right eigenvectors for a non-hermitian matrix, we define the left operator:
\begin{equation}
    \hat{\rho}_\eta^L = \hat{\rho}_S^{a, b, L} = \sum_{M, M^\prime}\delta_{M M^\prime}\ket{b, S, M^\prime}\bra{a, S, M}
    \label{eq:rho_eta_L}
\end{equation}
so that the inner product of the left and right operators can be defined as the trace of their matrix product:

\begin{equation}
\begin{split}
    &Tr(\hat{\rho}_\eta^L \hat{\rho}_{\eta^\prime}) = Tr[(\sum_{\mu, \mu^\prime} \delta_{\mu \mu^\prime} \ket{b, S, \mu^\prime} \bra{a, S, \mu})\\
    &(\sum_{M, M^\prime} \rho^{S^\prime}_{M M^\prime} \ket{a^\prime, S^\prime, M} \bra{b^\prime, S^\prime, M^\prime})]\\
    & = \delta_{\eta \eta^\prime} \sum_M \rho^S_{M M} = \delta_{\eta \eta^\prime}
    \label{eq:rho_eta_L_orth}
\end{split}
\end{equation}

The essence of the perturbation theory is to evaluate the transition among the basis operators $\hat{\rho}_\eta$ due to the perturbation. 

For dephasing, a kind of pure dissipative perturbation, the perturbed dynamics can be written as:

\begin{equation}
    \Delta \Dot{\hat{\rho}} = \sum_n \frac{\Gamma_\phi}{4}(\hat{\sigma}_n^z \hat{\rho} \hat{\sigma}_n^z - \hat{\rho}).
    \label{eq:eom_perturbation}
\end{equation}

For first-order perturbation theory, we substitute the steady-state density operator in Eq. (\ref{eq:rho_ss}) into the perturbed dynamics in Eq. (\ref{eq:eom_perturbation}) and take the projection on $\hat{\rho}_\eta$ by acting the left operator $\hat{\rho}_\eta^{L}$ on Eq. (\ref{eq:eom_perturbation}) and taking the trace:

\begin{equation}
\begin{split}
    &\Dot{R}_{\eta} = \frac{\Gamma_\phi}{4}\sum_n \sum_{\eta^\prime} Tr(\hat{\rho}_\eta^{L}\hat{\sigma}_n^z \hat{\rho}_{\eta^\prime} \hat{\sigma}_n^z - \hat{\rho}_\eta^{L} \hat{\rho}_{\eta^\prime})\\
    & = \frac{\Gamma_\phi}{4} \!\! \sum_{\eta^\prime} [O_{\eta \eta^\prime} - N \delta_{\eta \eta^\prime}]R_{\eta^\prime} = \!\! \sum_{\eta^\prime} C_{\eta \eta^\prime} R_{\eta^\prime} ,
\end{split}
\label{eq:C_eta_eta_prime_dephase}
\end{equation}
or in vector form:

\begin{equation}
    \Dot{\vec{R}} = \frac{\Gamma_\phi}{4} (\hat{O} - N \hat{I}) \vec{R} = \hat{C} \vec{R},
    \label{eq:vec_R_dot_C}
\end{equation}
where $\hat{O}$ is an $N_{ss} \times N_{ss}$ matrix, with matrix element defined as:
\begin{equation}
    O_{\eta \eta^\prime} \!\!= \!\!\! \sum_n \!\!\! \sum_{M, M^\prime} \!\!\! \rho_{M\!, M^\prime}^{S^\prime} \!\! \bra{a,\! S,\! M} \! \hat{\sigma}_n^z \! \ket{a^\prime\!,\! S^\prime\!,\! M} \!\! \bra{b^\prime\!,\! S^\prime\!,\! M^\prime}\!\hat{\sigma}_n^z\!\ket{b,\! S,\! M^\prime}
\end{equation}.

The matrix elements of $\hat{\sigma}_n^z$ in the angular momentum basis $\ket{a, S, M}$ can be evaluated using standard angular momentum techniques (for example, see Ref. \cite{edmonds1996angular}). 

For a pure $\Delta H$ type perturbation, the perturbed dynamics read:
\begin{equation}
    \Dot{\hat{\rho}} = -i [\Delta H \hat{\rho} - \hat{\rho} \Delta H].
    \label{eq:eom_perturbation_H}
\end{equation}

Similar to the dephasing case, using the first-order perturbation theory, we substitute the steady-state density operator in Eq. (\ref{eq:rho_ss}) into Eq. (\ref{eq:eom_perturbation_H}) and project to $\hat{\rho}_\eta$ by acting the left operator $\hat{\rho}_\eta^L$ on Eq. (\ref{eq:eom_perturbation_H}) and taking the trace:

\begin{equation}
    \dot{R}_\eta = -i \sum_{\eta^\prime} Tr(\hat{\rho}_\eta^L \Delta H \hat{\rho}_{\eta^\prime} - \hat{\rho}_\eta^L \hat{\rho}_{\eta^\prime} \Delta H) R_{\eta^\prime}
    \label{eq:dot_R_eta_H}
\end{equation}

Substitute the definition of $\hat{\rho}_{\eta}^L$ in Eq. (\ref{eq:rho_eta_L}) and $\hat{\rho}_{\eta^\prime}$ in Eq. (\ref{eq:rho_eta}) into Eq. (\ref{eq:dot_R_eta_H}) and lead to:

\begin{equation}
\begin{split}
    \dot{R}_\eta = \sum_{\eta^\prime} C_{\eta \eta^\prime} R_{\eta^\prime},
\end{split}
\label{eq:R_dot_C}
\end{equation}
where the matrix element of the coupling matrix is:
\begin{equation}
\begin{split}
     C_{\eta \eta^\prime} &= -i \sum_{M, M^\prime} \rho_{M M^\prime}^{S^\prime} \delta_{S S^\prime} (\delta_{bb^\prime}\bra{a, S, M^\prime}\Delta H \ket{a^\prime, S^\prime, M}\\
    & - \delta_{a a^\prime}\bra{b^\prime, S^\prime, M^\prime} \Delta H \ket{b, S, M})
\end{split}
\label{eq:C_eta_eta_prime_H}
\end{equation}

The matrix elements in Eq. (\ref{eq:C_eta_eta_prime_H}) can be evaluated using the standard angular momentum techniques in \cite{edmonds1996angular}.
\
\nocite{*}

\bibliography{ref}

\end{document}